\documentstyle[12pt]{article}

\def\be{\begin{equation}}
\def\ee{\end{equation}}
\def\ben{$$}
\def\een{$$}
\def\ba{\begin{array}{c}}
\def\ea{\end{array}}
\def\p{\partial}
\begin{document}

\titlepage
\vspace*{2cm}

 \begin{center}{\Large \bf
Non-Hermitian supersymmetry
\\
and
 singular, ${\cal PT}-$symmetrized oscillators
  }\end{center}

\vspace{10mm}

 \begin{center}
Miloslav Znojil

 \vspace{3mm}

\'{U}stav jadern\'e fyziky AV \v{C}R, 250 68 \v{R}e\v{z}, Czech
Republic\footnote{e-mail: znojil@ujf.cas.cz}

\end{center}

\vspace{5mm}

\section*{Abstract}

Hermitian supersymmetric partnership between singular potentials $
V(q)=q^2+{G}/{q^2}$ breaks down and can only be restored on
certain {\it ad hoc} subspaces [Das and Pernice, Nucl. Phys. B 561
(1999) 357]. We show that within extended, ${\cal PT}-$symmetric
quantum mechanics the supersymmetry between singular oscillators
can be completely re-established in a way which is continuous near
$G=0$ and leads to a new form of the bosonic creation and
annihilation operators.

\vspace{5mm}

PACS   03.65.Fd; 03.65.Ca; 03.65.Ge; 11.30.Pb; 12.90.Jv

\newpage

\section{Introduction \label{jedna}}

In the Witten's supersymmetric quantum mechanics \cite{Witten}, an
exceptionally important role is played by the linear harmonic
oscillator
 \ben
 H^{(LHO)}=p^2+q^2.
 \een
{\it A priori}, one would expect that inessential modifications of
$ H^{(LHO)}$ will exhibit nice properties as well. Unfortunately,
it is not so.  An elementary counterexample is due to Jevicki and
Rodriguez \cite{JR} who tried to combine $H^{(LHO)}$ with a
strongly singular supersymmetric partner
 \ben
 H^{(SHO)}=p^2+q^2+\frac{G}{q^2},
  \ \ \ \ \ \ \ G \neq 0.
 \een
They discovered that the expected supersymmetric correspondence
between their two models proves broken in a way attributed easily
to the strongly singular spike in the potential (cf., e.g.,
Section 12 of the review paper~\cite{Khare} for more details).

Recently, several people returned to this challenging problem
\cite{dopoc,Das,web,dopob,tata}. Independently, these authors have
found that a resolution of the Jevicki's and Rodriguez' puzzle
should be sought in a suitable regularization recipe. This
partially broadened the range of the Witten's supersymmetric
quantum mechanics. At the same time, the ambiguity of the choice
of the regularization itself remained a weak point of this
promising approach. For example, in the formulation of Das and
Pernice \cite{Das}, ``every distinct solution" (of the given
Schr\"{o}dinger equation) ``corresponds to a distinct
supersymmetrization". This means that the {\em superpotential may
cease to be state-independent}, with the partnership remaining
incomplete, {\em projected} on a mere subspace of solutions. The
subsequent re-formulations of this approach (say, in refs.
\cite{web,dopob}) weakened the latter disadvantage by narrowing
further the class of the regularized superpotentials. In
particular,  for $0 < G < 1$, the use of the continuous
superpotentials helps one to determine the spectrum via certain
nonlinear potential algebras (cf. \cite{dopob}) and/or via a
suitable limiting transition to the various new and still
``solvable" delta-function-type singular barriers (cf. also
refs.~\cite{tata}).

All these observations encouraged and inspired our present study.
We shall employ, in essence, the philosophy of our unpublished
preprint \cite{web}, the key idea of which may be further traced
back to the paper \cite{BG} on quartic anharmonicities in more
dimensions. There, Buslaev and Grecchi imagined that the
centrifugal barrier $G/q^2$ is an isolated pole of an analytic
potential in complex plane. Its most natural regularization is
analytic continuation mediated, say, by a {\em small complex
shift} of the coordinate axis $q \in I\!\!\!R \to r \in \,
l\!\!\!C$ such that, say,
 \be
 r = r(x)=x-i\,\varepsilon, \ \ \ \ \ \ x \in (-\infty,\infty).
 \label{regularization}
 \ee
Our text starts with a review of the properties of the eigenstates
of $H^{(SHO)}$ in both their centrifugal and regularized
interpretations (section \ref{tri}). Section \ref{dve} then
recollects a few basic facts about supersymmetry, amply
illustrated on $H^{(LHO)}$ and applied, subsequently, to the
regularized $H^{(SHO)}$. In a core of our message
(section~\ref{ctyricet}) the sets of wavefunctions are related by
non-Hermitian supersymmetry (SUSY).  Section~\ref{pet} finally
describes an interesting consequence in which a two-step
application of the SUSY mapping leads to the new concept of the
creation and annihilation operators.

\section{Singular oscillators \label{tri}}

\subsection{Centrifugal barrier \label{triv}}

Oscillator Hamiltonian $ H^{(LHO)}$ is easily generalized to more
dimensions $D = 2, 3, \ldots$. Fortunately, its partial
differential Schr\"{o}dinger equation
 \ben
\left ( -\triangle +|\vec{q}|^2
 \right ) \psi(\vec{q})
 = E\, \psi(\vec{q})
 \een
is superintegrable, i.e., separable in more ways
\cite{Winternitz}. In the spherical coordinates with $q =
|\vec{q}|$ it degenerates to the ordinary (so called radial)
differential equation
 \be
  H^{(\alpha)}
 \,\psi_{}^{}(q) = E_{}^{}(\alpha)
 \psi_{}^{}(q),
 \label{SE}
  \ee
 \ben
  H^{(\alpha)}= -
 \frac{d^2}{dq^2}
+
 \frac{\alpha^2-1/4}{q^2}
+ q^2,  \ \ \ \ \ \ \alpha=\alpha(\ell)=(D-2)/2 +\ell, \ \ \ \ \ \
\ell=0, 1, \ldots
 \een
defined on the half-line at any angular momentum. This explains
the exact solvability of the one-dimensional model $ H^{(SHO)}$
since its Schr\"{o}dinger equation differs
from eq.~(\ref{SE}) by the shift of $G=\alpha^2 -1/4$. In the
same vain, the original isotropic harmonic oscillator
and its smooth well
$V \sim |\vec{q}|^2$ may be complemented by any additional
singular central force $V' \sim \omega/|\vec{q}|^2$. Without any
loss of separability we re-define
 \ben
 \alpha =
 \alpha(\ell)= \sqrt{\omega+\left (
 \ell+\frac{D-2}{2}
 \right )^2}\ , \ \ \ \ \ \ \ell = 0, 1, \ldots
 \een
and the same solvable eq. (\ref{SE}) is to be considered.

\subsection{ ${\cal PT}$ symmetric solutions
\label{pion} }

The innocent-looking complex deformation $ q \to r(x)$ of
coordinates regularizes any centrifugal-like singularity $1/q^2$.
This modifies in fact the whole quantum mechanics in a way
advocated and made popular by Carl Bender et al \cite{BB}. Their
formalism works with non-Hermitian Hamiltonians which still
commute with the product of parity ${\cal P}$ and time reversal
${\cal T}$.  Such a type of a ``weakening" of the Hermiticity can
(though need not) support the real spectra and specifies an
extended, so called ${\cal PT}$ symmetric quantum mechanics
\cite{BBjmp}, intensively studied in the mathematically oriented
contemporary literature~\cite{Dorey}. The related enhanced
interest in analyticity has already proved useful in some
applications, {\it inter alii}, in the context of perturbation
theory \cite{Caliceti}, field theory~\cite{BM} and, last but not
least, supersymmetric quantum mechanics~\cite{Andrianov}.  In
principle, the ordinary Sturm-Liouville theory must be adapted to
the new situation \cite{Sturm}, the norms have to be replaced by
the pseudo-norms \cite{whatime} etc. All these technical aspect of
${\cal PT}$ symmetry may, fortunately, be skipped here as
inessential since the regularization represented by eq.
(\ref{regularization}) is all we shall need in what follows.  In
this sense, the present application of the ${\cal PT}$ symmetric
formalism to the spiked harmonic oscillators will shift the line
of coordinates and recall the ${\cal PT}$ symmetric analytic
solution of the resulting Schr\"{o}dinger equation~(\ref{SE}) as
described in ref. \cite{ptho}. For any non-negative $\alpha\geq 0$
which, for technical reasons, is not equal to an integer, $\alpha
\neq 0, 1, 2, \ldots$, we get the spectrum
  \be
  E_{}^{}
  = E_{N}^{(\varrho)} =4N+2\varrho+2,
\ \ \ \ \ \ \ \
 \varrho= -Q \cdot
\alpha
  \label{key}
  \ee
numbered by the integers $N = 0, 1, \ldots$ and by the so called
quasi-parity $Q = \pm 1$. The related wavefunctions are
represented in terms of the Laguerre polynomials,
 \be
 \psi_{ }^{}(r)
= \langle r | N, \varrho \rangle =
\frac{N!}{\Gamma(N+\varrho+1)}\cdot
 r^{\varrho+1/2} \exp(-r^2/2) \cdot L_N^{(\varrho)}(r^2).
 \label{keyf}
  \ee
The quasi-parity $Q$ is defined in such a way that it coincides
with the ordinary spatial parity $P$ in the limit $\varepsilon
\to 0$. This convention puts the quasi-even level
$E^{(-\alpha)}$ with the dominating threshold behaviour $\psi(r)
\sim r^{1/2-\alpha}$ lower than its quasi-odd complement
$E^{(+\alpha)}$ with the dominated threshold behaviour $\psi(r)
\sim r^{1/2+\alpha}$ at any fixed $N$.  In this way, the
Hermitian limit $\varepsilon \to 0$ leads to the necessity of
elimination of the former, quasi-even solutions as unphysical
(i.e., quadratically non-integrable) whenever~$\alpha \geq 1$.

Our bound states degenerate to the well known eigenstates of the
linear harmonic oscillator at $\alpha = 1/2$. Marginally, let us
note that in the other regularization schemes the correspondence
between $P$ and $Q$ may be different.  The ambiguity is due to the
strongly singular character of the core $1/q^2$.  Thus, in the
matching recipe of section~3 in ref. \cite{Das} for example, Das
and Pernice recommend an exclusive use of $Q=+1$.  The spatial
parity $P=\pm 1$ is then introduced in non-analytic manner. A
continuous extension of this recipe to the regular case with
$\alpha=1/2$ is, therefore, impossible.

\section{Supersymmetry
\label{dve}}

For the linear harmonic oscillator
the Schr\"{o}dinger's factorization method \cite{Hull} in
application to the Hamiltonian
$H^{(LHO)}$
offers a nice illustration of the essence of the supersymmetric
quantum mechanics.

\subsection{Example
\label{dver}}

Let us remind the reader that
$H^{(LHO)} =A\cdot B - 1 = B \cdot A + 1$ with $A = q+ip$ and $B = q-ip$.
This
enables us to define a pair of the partner Hamiltonians, viz,
 the ``left" $H_{(L)}=H^{(LHO)}-1=B\cdot A$ and the
``right" $H_{(R)}=H^{(LHO)}+1=A\cdot B$.
One can easily verify that
their factorization implies that the
so called ``super-Hamiltonian" and two ``supercharges"
 \ben
 {
 \cal H}= \left [ \begin{array}{cc} H_{(L)}&0\\ 0&H_{(R)}
 \ea
 \right ]
, \ \ \ \ \ \
 {
 \cal Q}=\left [
 \begin{array}{cc} 0&0\\ A^{}&0
 \ea
 \right ],
 \ \ \ \ \ \
\tilde{\cal Q}=\left [
 \begin{array}{cc}
0& B^{}
\\
0&0 \ea \right ]\
 \een
generate a representation of Lie superalgebra sl(1/1),
 \ben
 \{ {\cal Q},\tilde{\cal Q}
\}={\cal H} , \ \ \ \ \ \ \{ {\cal Q},{\cal Q} \}= \{ \tilde{\cal
Q},\tilde{\cal Q} \}=0, \ \ \ \ \ \ \ \ [ {\cal H},{\cal Q} ]=[
{\cal H},\tilde{\cal Q} ]=0.
  \een
In this language the creation,
annihilation and occupation-number operators
are easily defined for fermions,
 \ben
 {
 \cal F}^\dagger=\left [
 \begin{array}{cc} 0&0\\ 1^{}&0
 \ea
 \right ],
 \ \ \ \ \ \
{\cal F}=\left [
 \begin{array}{cc}
0& 1^{}
\\
0&0 \ea \right ],
 \ \ \ \ \ \
{\cal N}_{\cal F}=\left [
 \begin{array}{cc}
0& 0^{}
\\
0&1 \ea \right ]\ .
 \een
In the bosonic sector, the Fock-space structure is, generically,
more complicated (cf., e.g., Sections 2 and 8 in the review
\cite{Khare}). It only becomes simplified for the present
harmonic-oscillator example where the creation and/or annihilation
of a boson remains mediated by the first-order differential
operators $ {\bf a}^\dagger\sim B$ and $ {\bf a}\sim A$, respectively.
This enables us to
work with the factorized supercharges ${\cal Q} \sim {\bf
a}{\cal F}^\dagger$ and $\tilde{\cal Q} \sim {\bf a}^\dagger{\cal F}$
and with the following elementary vacuum,
 \ben
 \langle q |0\rangle = \left [
 \ba
 \exp(-q^2/2)/\sqrt{\pi} \\
 0
 \ea
 \right ], \ \ \ \ \
 {\cal Q}\,|0\rangle =
 \tilde{\cal Q}\,|0\rangle = 0.
 \een
We may summarize that in this model the supersymmetry between
bosons and fermions is unbroken and explicitly represented in
Fock space (cf. \cite{Khare}, p. 283).

\subsection{Superpotentials \label{petr} }

Even beyond the above elementary
harmonic-oscillator illustration, all the
{\em Hermitian} supersymmetric quantum mechanics
is based on the
Schr\"{o}dinger's factorization
of the Hamiltonians
(cf. review \cite{Khare}).
These constructions {\em start} from the
so called superpotential $W$ and from the doublet of the
explicitly defined
operators $A=\p_q+W$ and $B=-\p_q+W$. This leads to the
supersymmetry as described in subsection \ref{dver} and to the two
partner Hamiltonian operators which are different from each other
in general,
 \be
 H_{(L)}=B\cdot A=\hat{p}^2+W^2-W'
 , \ \ \ \ \ \ \ \
 H_{(R)}=A \cdot B=\hat{p}^2+W^2+W'.
 \label{partner}
 \ee
When we return to our regularization recipe
 $q \to
r(x)=x-i\,\varepsilon$, we may re-interpret
our ${\cal PT}$ symmetric Schr\"{o}dinger equation (\ref{SE})
 as a regular complex equation
on the real line of $x$. It is then easy to introduce the
superpotential we need,
 \be
 W_{}^{(\gamma)}(r) = - \frac{\p_r
 \langle r | 0, \gamma \rangle
}{
 \langle r | 0, \gamma \rangle
}=
r-\frac{\gamma+1/2}{r}\, \ \ \ \ \ \ \ r = r(x).
 \label{N}
 \ee
This function
 is regular at all the real $x$. In the other words, we
start from the choice of a real parameter
$\gamma $ and from the knowledge of the related superpotential
(\ref{N}) and {\em define} the pair (\ref{partner}) afterwards.
In this step we already get an interesting pattern which is
summarized in Table~1. The supersymmetric recipe gives the
$\gamma-$numbered partner Hamiltonians in the explicit and compact
harmonic oscillator form,
 \be
 {H}_{(L)}^{(\gamma)} = {H}_{}^{(\alpha)} -2\gamma-2,
 \ \ \ \ \ \
 {H}_{(R)}^{(\gamma)} = {H}_{}^{(\beta)} -2\gamma, \ \ \ \
 \ \ \ {\alpha}=|\gamma|, \ \ \ \
 \ \ \ \beta=|\gamma+1|\ .
 \label{Mtt}
 \ee
In the light of eqs. (\ref{key}) and (\ref{keyf}) the energies and
wavefunctions are given by the elementary formulae.

\subsection{Spectra}

We have to distinguish between the three intervals of $\gamma$
because the Hamiltonians depend on the absolute values
$\alpha=|\gamma|$ and $\beta=|\gamma+1|$. This implies that for
the fixed parameter $\gamma$ and at any principal quantum number
$N$, we have the four different wavefunctions distinguished by the
subscripts $_{(L)}$ and $_{(R)}$  [{or} arguments $\alpha$ and
$\beta$, respectively] and by the two quasi-parities $Q=\pm 1$.
This gives the four kets
 \ben
 |N,-\alpha \rangle, \ \ \ \ \
 |N,-\beta \rangle, \ \ \ \ \
 |N,+\alpha \rangle, \ \ \ \ \
 |N,+\beta \rangle
 \een
corresponding to the respective energies
 \be
 E_{(L)}^{(-\alpha)} \leq
 E_{(R)}^{(-\beta)} \leq
 E_{(L)}^{(+\alpha)} \leq
 E_{(R)}^{(+\beta)}.
  \label{ordered}
 \ee
At any $N = 0, 1, \ldots$ these energies are ordered in a
$\gamma-$independent manner. This is illustrated in Figure~1 which
displays the $\gamma-$dependence of the low lying spectrum for our
supersymmetrized system (\ref{Mtt}). In the Figure, an interplay
between the ordering and degeneracy is made visible by the
infinitesimal $\eta \to 0$ shifts of the energies,
 \be
L(+N) = E_{(L)}^{(-\alpha)}-2\,\eta, \ \ \ \ \ \ L(-N) =
E_{(L)}^{(+\alpha)}+\eta,
 \label{shiftsa}
 \ee
 \be  R(+N) = E_{(R)}^{(-\beta)}-\eta,
 \ \ \ \ \
R(-N) =  E_{(R)}^{(+\beta)}+2\,\eta.
 \label{shiftsb}
 \ee
With all $N$ included, all the energy levels become doubly
degenerate, with the single exception of~$E = 0$.  This is, as
we know, characteristic for the supersymmetric
quantum-mechanical models where the so called Witten's
index~\cite{Wittenind} does not vanish.

We may notice that the level $E=0$ coincides with the ground
state energy $L(+0)$ if and only if $ \gamma < 0$. On the
opposite half-line of $\gamma > 0$, the vanishing energy
$L(-0)=0$ acquires the negative quasi-parity while the
quasi-even and doubly degenerate ground-state energy becomes
strictly negative, $L(+0)=R(+0)<0$. The latter feature of our
present consequent non-Hermitian supersymmetrization is in a
sharp contrast to the strict non-degeneracy of the ground states
in the Hermitian cases.

\section{Wavefunctions \label{ctyricet}}

\subsection{Supercharges}

The coincidence of the strengths of
the spike in our two partner Hamiltonians
(\ref{Mtt})
 is possible but fairly exceptional. Indeed,
postulating that $\alpha =\alpha_e=\beta=\beta_e$, the related
parameter $\gamma_e$ becomes specified by the algebraic equation
$|\gamma_e|=|\gamma_e+1|$. Its solution is unique, $\gamma_e =
-1/2$, and makes our superpotential (\ref{N}) regular. In Figure~1
we may check that such a choice gives the equidistant LHO
spectrum.

All the other admissible (i.e., non-integer and real) values
of~$\gamma$ lead to the singular supercharge components
 \be
 A^{(\gamma)}=\p_r+W^{(\gamma)}, \ \ \ \ \ \
 B^{(\gamma)}=-\p_r+W^{(\gamma)}, \ \ \ \ \ \ \gamma \neq 0, \pm
 1, \ldots .
 \ee
They act on our (normalized, spiked and ${\cal PT}-$symmetric)
Laguerre-polynomial states
 \ben
 \langle r | N, -Q\cdot \alpha \rangle \equiv {\cal L}^{(-Q\cdot
 \alpha)}_{N}(r)
 \een
in an extremely transparent and compact manner,
 \be
 \label{rulesa}
 A^{(\gamma)}\, {\cal
 L}^{(\gamma)}_{N+1}{}=c_1(N,\gamma)\, {\cal
 L}^{(\gamma+1)}_{N}{}, \ \ \ \ \ \ \ \ \
\ c_1(N,\gamma)=-2\sqrt{N+1}; \ee
 \be
 \label{rulesb}
 B^{(\gamma)}\, {\cal
 L}^{(\gamma+1)}_{N}{}=c_2(N,\gamma)\, {\cal
 L}^{(\gamma)}_{N+1}{}, \ \ \ \ \ \ \ \ \
\ c_2(N,\gamma)=-2\sqrt{N+1}; \ee
 \be A^{(\gamma)}\, {\cal
 L}^{(-\gamma)}_{N}{}=c_3(N,\gamma)\, {\cal
 L}^{(-\gamma-1)}_{N}{}, \ \ \ \ \ \ \ \ \
\ c_3(N,\gamma)=2\sqrt{N-\gamma};
 \label{rulesc}
 \ee
  \be
 \label{rulesd}
 B^{(\gamma)}\, {\cal
 L}^{(-\gamma-1)}_{N}{}=c_4(N,\gamma)\, {\cal
 L}^{(-\gamma)}_{N}{} \ \ \ \ \ \ \ \ \
\ c_4(N,\gamma)=2\sqrt{N-\gamma}.
 \ee
This is our main formula. Its first two lines prove sufficient
to define the well known one-dimensional annihilation and
creation at $\alpha = 1/2$. The latter two lines find their
application at any $\alpha \neq 1/2$.
For the slightly non-LHO choice of $\gamma = 2/5$ this is
illustrated in Table~2.

We see an explicit $\gamma-$dependence in $c_3$ and $c_4$. These
coefficients would vanish (and mimic a ``false-vacuum") at any
integer $\gamma$.  This is an additional, algebraic reason for our
elimination of $\gamma=$integer, complementing the analytic
pathology of these points (viz., an unavoided level crossing) as
observed previously in ref. \cite{ptho}.

\subsection{Hermitian limit}

The ``natural" domain of parameter $\gamma \notin Z\!\!\!Z$ in our
superpotential (\ref{N}) is real line, $\gamma \in
(-\infty,\infty)$.  In the Hermitian limit $\varepsilon \to 0$,
this domain has to be split in the five separate subdomains, viz.,
the ``far left" ${\cal D}_{(fl)}=(-\infty,-2)$, the ``near left"
${\cal D}_{(nl)}=(-2,-1)$, the above-mentionned ``centre" ${\cal
D}_{(c)}=(-1,0)$, the ``near right" ${\cal D}_{(nr)}=(0,1)$ and
the ``far right" ${\cal D}_{(fr)}=(1,\infty)$.

In the leftmost and rightmost intervals ${\cal D}_{(fl)}$ and
${\cal D}_{(fr)}$ the respective quasi-even ${\cal PT}$ symmetric
doublets $[L(+N+1),R(+N)]$ and $[L(+N),R(+N)]$ become
non-normalizable and disappear from our horizon completely. Up to
that expected reduction, the limit $\varepsilon \to 0$ does not
change the original ${\cal PT}$ symmetric spectrum. In both the
latter domains, the SHO supersymmetry is established in a more or
less textbook form.

Within the neighboring further two intervals ${\cal D}_{(nl)}$ and
${\cal D}_{(nr)}$, the supersymmetry would be completely destroyed
by the survival of the respective quasi-even normalizable
solutions ${\cal L}_N^{(-\beta)}(r)$ or ${\cal
L}_N^{(-\alpha)}(r)$.  Similar ``redundant" sets of solutions have
already been reported as causing serious difficulties \cite{JR},
\cite{Khare}, \cite{Das}.  In the light of our present results the
supersymmetic correspondence between the Hamiltonians
$H^{(\alpha)}$ and $H^{(\beta)}$ may be again fully restored
within the latter two domains.  One even need not resort to any
sophisticated reasons because the necessary elimination of the
supersymmetry breaking wavefunctions can simply be performed by
using an auxiliary boudary condition in the origin,
 \be
 \lim_{r \to 0} \frac{\psi(r)}{\sqrt{r}} = 0.
 \label{bc}
 \ee
Fortunately, this condition coincides with the standard physical
constraint for the radial wavefunctions in more dimensions
\cite{pra}.  Hence, we may easily interpret this constraint as a
mere return to the standard supersymmetry {\em without}
singularities. Indeed, our superpotential $W$ has no singularities
within the range $r \in (0,  \infty)$ of the radial coordinate
at~$D > 1$.

A remarkable situation is encountered in ${\cal D}_{(fl)}$ and
${\cal D}_{(nl)}$ where our $\varepsilon \to 0$ supersymmetry
could be characterized, conventionally, as broken (cf. p. 285 in
\cite{Khare}).  Its more-dimensional re-interpretation becomes
necessary in ${\cal D}_{(nl)}$ again.

One of our most amazing conclusions concerns the ``central"
interval ${\cal D}_{(c)}$ where our ${\cal PT}$ symmetric
regularization can very easily be removed and the picture provided
by Figure~1 applies in the Hermitian case without any changes.

We may conclude that the ${\cal PT}$ symmetric formalism leads to
the Hermitian limits which exhibit the correct supersymmetric
correspondence between Hamiltonians (\ref{Mtt}) at almost all the
parameters $\gamma$.  During the limiting transition $\varepsilon
\to 0$ the Hermitian spectra may be reduced but the supersymmetry
survives.  Roughly speaking, we re-established a full
supersymmetry between the ``left" and ``right" SHO systems simply
via their $s-$wave re-interpretation.  This conclusion is
summarized in Table~3.

\section{Innovated annihilation and creation
\label{pet} }

\subsection{Definition}

At $\gamma=-1/2$ we encounter the ``degenerate" (and, in the
present context, utterly exceptional) textbook LHO
pattern
 \ben
A^{(-1/2)}\,
\cdot  {\cal L}_{N-1}^{(1/2)}(q)
\sim
   {\cal L}_{N-1}^{(-1/2)}(q) ,
 \ \ \ \ \ \ \ \ \
A^{(-1/2)}\,
\cdot  {\cal L}_{N}^{(-1/2)}(q)
 = - \sqrt{2N}\,  {\cal L}_{N-1}^{(1/2)}(q)
 \een
 \ben
B^{(-1/2)}\,
{\cal L}_{N-1}^{(-1/2)}(q)
\sim  {\cal L}_{N-1}^{(1/2)}(q) , \ \ \ \ \
\ \ \ \
B^{(-1/2)}\,
 {\cal L}_{N-1}^{(1/2)}(q)
 \sim  {\cal L}_{N}^{(-1/2)}(q).
 \een
The second half of Table 4 (which, as a whole, will be needed
later) offers a remarkable alternative. Indeed, via the
non-Hermitian detour and limit $\varepsilon \to 0$, another
explicit annihilation pattern is obtained for the same $s-$wave
oscillator. The new SUSY mapping would start from the Hamiltonian
${H_{(L)}=H^{(1/2)}-3}$ giving its ${\cal PT}$ symmetrically
regularized non-Hermitian partner $H_{(R)}=H^{(3/2)}-1$. In the
subsequent step (and  in a way indicated, up to the shifts which
are different, in the first half of Table~4), the similar SUSY
partnership of the re-shifted $ H_{({L})}=H^{(3/2)}+1$ would
return us to the re-shifted original ${H_{({R})}=H^{(1/2)}+3}$.

All these examples indicate that the annihilation operators and
their creation partners can be introduced in the factorized,
second-order differential form
 \be
A^{(-\gamma-1)} \cdot
A^{(\gamma)}
=
A^{(\gamma-1)}
\cdot
A^{(-\gamma)}
=
{\bf A}(\alpha),
\label{operatorsa}
 \ee
 \be
B^{(-\gamma)}
\cdot
B^{(\gamma-1)}=
B^{(\gamma)}
\cdot
B^{(-\gamma-1)}=
{\bf A}^\dagger (\alpha).
\label{operatorsb}
 \ee
Once we start from $\alpha=3/2$ this observation may be
illustrated by the two alternative superpositions of the action of
the supercharges $A^{(\gamma)}$ as displayed in Tables~4 and~5. In
the former one the $\gamma=-3/2$ ${\cal PT}$ supersymmetry between
${H_{(L)}=H^{(3/2)}+1}$ and $H_{(R)}=H^{(1/2)}+3$ is followed by
the  $\gamma=1/2$ correspondence between the doublet
${H_{(\tilde{L})}}=H^{(1/2)}-3$ and
${H_{(\tilde{R})}=H^{(3/2)}-1}$. As a net result we obtain the
appropriate generalization of the annihilation pattern for the
harmonic oscillator in $p-$wave. Table~5 offers an alternative
path again.

\subsection{Action}

At a general $ \alpha \neq 0, 1, 2, \ldots$, the operators
(\ref{operatorsa}) and (\ref{operatorsb}) enable us to move along
the spectrum of any spiked harmonic oscillator Hamiltonian
$H^{(\alpha)}$. We get the elementary and transparent action on
all the solutions, \ben
 {\bf A}(\alpha) \cdot
 {\cal L}^{(\gamma)}_{N+1}{}=c_5(N,\gamma)\,
 {\cal L}^{(\gamma)}_{N}{},
 \een
 \ben
{\bf A}^\dagger(\alpha) \cdot
 {\cal L}^{(\gamma)}_{N}{}=c_5(N,\gamma)\,
 {\cal L}^{(\gamma)}_{N+1}{},
 \een
 \ben
c_5(N,\gamma)=-4\sqrt{(N+1)(N+\gamma+1)}, \ \ \ \ \ \ \ \
\gamma=\pm \alpha.
 \een
We achieved a unified description of the spiked harmonic
oscillators $H^{(\alpha)}$ within the ${\cal PT}$ symetric
framework.

\begin{itemize}

\item
The ${\cal PT}$ supersymmetric partnership is mediated by the
first-order differential operators $A^{(\gamma)}$ and $B^{(\gamma)}$.

\item
At any non-integer $\alpha>0$ in the Hamiltonian $H^{(\alpha)}$
the role of the creation and annihilation operators is played by
the $\alpha-$dependent and $\gamma-$preserving differential
operators
 ${\bf A}^\dagger(\alpha)$ and
${\bf A}(\alpha)$ of the second order.

\end{itemize}

 \noindent
The ${\cal PT}$ supersymmetric partners coincide solely in the
regular case. Its traditional creation and annihilation operators
${\bf a}^\dagger \sim B{(-1/2)}$ and ${\bf a} \sim A{(-1/2)}$ {\em
change} the quasi-parity. This feature is not transferable to any
non-equidistant spectrum with $\gamma \neq -1/2$.

Our ``natural" operators of creation ${\bf A}^\dagger(\alpha)$ and
annihilation ${\bf A}(\alpha)$ are smooth near $\alpha = 1/2$.
Their marginal (though practically relevant) merit lies in their
reducibility to their regular first-order differential
representation
 \ben
 {\bf A}(\alpha) \cdot
 {\cal L}^{(\gamma)}_{N}{}=
  (2r\p_r+2\,r^2-4N-2\gamma-1)\cdot
 {\cal L}^{(\gamma)}_{N}{},
 \een
 \ben
 {\bf A}^\dagger(\alpha) \cdot
 {\cal L}^{(\gamma)}_{N}{}=
  (-2r\p_r+2\,r^2-4N-2\gamma-3)\cdot
 {\cal L}^{(\gamma)}_{N}{}
 \een
which is, of course, state-dependent. The further change of
variables $r \to y$ such that $r=\exp 2y$ gives a simpler
differentiation $2r\p_r \to \p_y$ and the Morse Hamiltonians with
${\cal PT}$ symmetry~\cite{Morse}. This indicates that the Morse
potentials would also deserve more attention in the supersymmetric
context.

\section{Summary}

In their inspiring letter \cite{JR} Jevicki and Rodriguez
emphasized that the supersymmetric partnership cannot be
postulated between $H_{(L)}=H^{(LHO)}-3$ (with energies
$E_0^{(+1/2)}=-2$, $E_0^{(-1/2)}=0$, $E_1^{(+1/2)}=2$,
$E_1^{(-1/2)}=4$ etc) and $H_{(R)}=p^2+q^2+2/q^2-1$ (with the
different set of the levels $E_0^{(-3/2)}=4$, $E_1^{(-3/2)}=8$
etc). We have seen that the puzzle is resolved when we treat both
operators as $s-$wave Hamiltonians. This reduces the ``left"
spectrum to the new set ($E_0^{(-1/2)}=0$, $E_1^{(-1/2)}=4$ etc)
and the supersymmetry is restored.

The problem recurred when Das and Pernice \cite{Das} did not find
any analogy between $\alpha = 1/2$ (smooth, LHO) and $\alpha \neq
1/2$ (spiked, singular SHO). In their method, different approaches
were required as long as a few {\it a priori}
supersymmetry-supporting requirements (e.g., of the existence of a
non-degenerate ground state at $E=0$) were  postulated. As a
consequence, the supersymmetry of ref. \cite{Das} did not apply to
the pairs of operators but rather to their {\it ad hoc}
projections which were not always clearly specified.

The non-analytic regularization of ref. \cite{Das} was also
unnecessarily complicated. For example, the regularization giving
the even quasi-parity $Q=-1$ (as mentioned at the end of the
subsection \ref{pion} above) was only chosen consequently at the
even spatial parity $P=-1$. For $P=+1$ one uses  $Q=-1$  at
$\gamma > 0$ for the ``left" $H_{(L)}$ and at $\gamma <-1$ for the
``right" $H_{(R)}$. For the other $\gamma$ it was necessary to use
the quasi-even solutions with $Q=+1$, anyhow.

These problems have been resolved in the present alternative
approach. We have shown that a key to the problem lies in the
suitable non-Hermitian regularization of the singular
superpotentials.  Although this merely circumvents the problem
with the singularity in $H^{(SHO)}$,  we need not really remove
the regularization in the majority of phenomenological and
supersymmetric applications. It suffices, mostly, to stay suitably
(though not too much) close to the limit, keeping the
Schr\"{o}dinger equations comfortably non-singular. Moreover,
there exist serious mathematical reasons why one should avoid the
removal of the regularization whenever possible. In one dimension,
the $1/q^2$ barrier  {\em always} separates the real line,
strictly speaking, into two non-communicating halves~\cite{Reed}.

In our paper we have advocated the use of the ${\cal PT}$
symmetric regularization (\ref{regularization}) which exihibits
several specific merits. First of all, it ``supersymmetrizes" the
pairs of Hamiltonians $H^{(SHO)}$ for all the couplings
$G=\alpha^2-1/4$ for which $\alpha=|\gamma|$ is not an integer,
$\gamma \notin Z\!\!\!Z$. Secondly, all the formulae degenerate to
the well known harmonic-oscillator supersymmetry at $\gamma =
-1/2$.  Thirdly, the limiting transition $\varepsilon \to 0$
proves smooth at all the neighboring $\gamma \in (-1,0)$.  This
enabled us to generalize the LHO model to all the SHO doublets
$H_{(L)}^{(\alpha)}$ and $H_{(R)}^{(\beta)}$ with $\alpha =
|\gamma|$ and $\beta = |\gamma+1|$.

\section*{Acknowledgement}

Work partially supported by the grant Nr. A 1048004 of GA AS CR.

\newpage

\section*{Figure captions}

Figure 1. The $\gamma-$dependence of the SHO spectrum generated by
superpotential~(\ref{N}).


\section*{Tables}


Table 1.

 \noindent
${\cal PT}$ supersymmetry
of harmonic oscillators at non-integer $\gamma$.

 $$
 \begin{array}{||c||c|c|c||}
 \hline \hline
  {\rm the\  range\ of\ }\gamma
  &(- \infty,-1)& (-1,0)&
  (0,\infty)\\
  \hline
  \hline
 {\rm parameters}&&&\\
   {\rm  }\alpha = |\gamma| > 0 &-\gamma&-\gamma&\gamma\\
 \beta=|\gamma+1| > 0&\alpha-1&1-\alpha&\alpha+1\\
  &&&\\
  \hline
  \hline
 {\rm Hamiltonians}&&&\\
  H_{(L)}
  &
  H^{(\alpha)}+  2\beta
  &
  H^{(\alpha)} -2\beta
  &
  H^{(\alpha)} -2\beta
  \\
  H_{(R)}
  &
  H^{(\beta)}  +2\alpha
  &
  H^{(\beta)}+2\alpha
  &
  H^{(\beta)}-2\alpha
  \\
  &&&\\
 \hline
  \hline
 {\rm energies}&&&\\
 E_{(L)}^{(\beta)}
   &4N+4\alpha
  &4N+4&4N+4
  \\
E_{(L)}^{(\alpha)}
   &4N+4\alpha&4N+4\alpha
  &4N
  \\
 E_{(L)}^{(-\beta)}
    &4N+4
  &4N+4\alpha&4N-4\alpha
  \\
E_{(L)}^{(-\alpha)}
  &4N&4N
  &4N-4\alpha
  \\
  &&&\\
  \hline
 \hline \ea $$

\newpage

 Table 2.

\noindent
The action of $A^{(\gamma)}$
near LHO, at $\gamma=-1/2+1/10=-2/5$

$$
\begin{array}{||c|ccc||}
\hline \hline
E_{(L)}=
E_{(R)}
 &|N_{(L)}\rangle&
{\longrightarrow} &|N_{(R)}\rangle\\
 \hline
\hline \vdots&\vdots&&\vdots
 \\
 8&{\cal L}^{(-2/5)}_{2}{}&\rightarrow&{\cal L}^{(3/5)}_{1}{}
 \\
 5.6&{\cal L}^{(2/5)}_{1}{}&\rightarrow&{\cal L}^{(-3/5)}_{1}{}
 \\
 4&{\cal L}^{(-2/5)}_{1}{}&\rightarrow&{\cal L}^{(3/5)}_{0}{}
 \\
 1.6&{\cal L}^{(2/5)}_{0}{}&\rightarrow&{\cal L}^{(-3/5)}_{0}{}
 \\
 0&{\cal L}^{(-2/5)}_{0}{}&\rightarrow&{0}{}
 \\
 \hline \hline
\ea $$

\newpage

Table 3.

 \noindent
Hermitian limit $\varepsilon \to 0$ in  Figure~1 and
supersymmetric correspondence between the spiked harmonic
oscillators (\ref{Mtt}).

 $$
 \begin{array}{||c||c|c|c|c|c||}
 \hline \hline
{\rm domain}
  & (fl)
  & (nl)
& (c) & (nr)
 &
  (fr)\\
\hline
  \gamma
  & (- \infty,-2)
  & (- 2,-1)
& (-1,0) & (0,1)
 &
  (1,\infty)\\
 \triangle^{1)} & 0^{2)}&0^{2)}&1^{3)}&1^{4)}&1^{4)}\\ E_{(L)}\
^{5)} & L(-N) & L(-N)
 &L(+N+1)^{6)} &L(-N-1)
  &L(-N-1)
 \\
{{\cal L}_N^{(-\alpha)}(r)}^{7)} &
 {\rm absent}^{8)}&
 {\rm absent}^{8)}&
 {\rm present} &
 {\rm dropped}^{9)}&
 {\rm absent}^{8)}
\\
{{\cal L}_N^{(-\beta)}(r)}^{10)} &
 {\rm absent}^{8)}&
 {\rm dropped}^{9)}&
 {\rm present} &
 {\rm absent}^{8)}&
 {\rm absent}^{8)}
\\
  \hline
{\rm SUSY} & {\rm broken}& {\rm broken}& {\rm unbroken}& {\rm
unbroken}& {\rm unbroken}\\ \hline
 \hline
\multicolumn{6}{l}{\rm footnotes}\\ \hline
 \multicolumn{6}{l}{
\mbox{}^{1)}
 {\rm Witten's\ index\ \cite{Wittenind} }
}\\
 \multicolumn{6}{l}{
\mbox{}^{2)}
 {\rm degenerate \ ground\ state\ at\  positive\ energy\ }
L(-0) =R(-0) =4\alpha }\\
 \multicolumn{6}{l}{
\mbox{}^{3)}
 {\rm nondegenerate\ ground\ state\ at\  energy\ }
L(+0)=0 }\\
 \multicolumn{6}{l}{
\mbox{}^{4)}
 {\rm nondegenerate\ ground\ state\ at\  energy\ }
L(-0)=0 }\\
 \multicolumn{6}{l}{
\mbox{}^{5)}
 {\rm supersymmetric\ partner\  of\ } E_{(R)}= R(-N)
}\\
 \multicolumn{6}{l}{
\mbox{}^{6)}
 {\rm
the\ second\ series\ has\   }
 E_{(L)}'= E_{(R)}' =
 {L(-N)}
 = {R(+N)
 }
}\\
 \multicolumn{6}{l}{
\mbox{}^{7)}
 {\rm quasi-even\ state\  with \ energy \ } L(+N)
}\\
 \multicolumn{6}{l}{
\mbox{}^{8)}
 {\rm not\ integrable }
}\\
 \multicolumn{6}{l}{
\mbox{}^{9)}
 {\rm  eliminated\ using\ an\ auxiliary
 \  boundary\ condition\ in\ the\ origin }
}\\
 \multicolumn{6}{l}{
\mbox{}^{10)}
 {\rm quasi-even\ state\  with \ energy \ } R(+N)
}\\ \hline \hline \ea $$

\newpage

 Table 4.

\noindent Singular Hamiltonian
$H^{(3/2)}=p^2+(x-i\varepsilon)^2+2/(x-i\varepsilon)^2$ and
annihilation operator as a double supersymmetric mapping with
initial $\gamma = -3/2$.

$$
\begin{array}{||c|ccccc|c||}
\hline \hline
E_{(L)}=E_{(R)}
 &|N_{(L)}\rangle& {\stackrel{A^{(-3/2)}
}{\longrightarrow}} &|N_{(R)}\rangle= |N_{(\tilde{L})}\rangle&
{\stackrel{A^{(1/2)} }{\longrightarrow}} &|N_{(\tilde{R})}\rangle&
E_{(\tilde{L})}=E_{(\tilde{R})}
\\
 \hline
\hline \vdots&\vdots&&\vdots&&\vdots&\vdots
\\
 14&{\cal L}^{(3/2)}_{2}{}&\rightarrow&{\cal L}^{(1/2)}_{2}{}&
\rightarrow&{\cal L}^{(3/2)}_{1}{}&8
 \\
 12&{\cal L}^{(-3/2)}_{3}{}&\rightarrow&{\cal L}^{(-1/2)}_{2}{}&
\rightarrow&{\cal L}^{(-3/2)}_{2}{}&6
 \\
 10&{\cal L}^{(3/2)}_{1}{}&\rightarrow&{\cal L}^{(1/2)}_{1}{}&
\rightarrow&{\cal L}^{(3/2)}_{0}{}&4
 \\
 8&{\cal L}^{(-3/2)}_{2}{}&\rightarrow&{\cal L}^{(-1/2)}_{1}{}&
\rightarrow&{\cal L}^{(-3/2)}_{1}{}&2
 \\
 6&{\cal L}^{(3/2)}_{0}{}&\rightarrow&{\cal L}^{(1/2)}_{0}{}
&\rightarrow&0&0
 \\
 4&{\cal L}^{(-3/2)}_{1}{}&\rightarrow&{\cal L}^{(-1/2)}_{0}{}&
 \rightarrow&{\cal L}^{(-3/2)}_{0}{}&-2
 \\
2&-&&-&&-&-4\\
 0&{\cal L}^{(-3/2)}_{0}{}&\rightarrow&{0}{}&
 \rightarrow&-&-6
 \\
 \hline \hline
\ea $$

\newpage

 Table 5.

\noindent Same as Table 4 with $\gamma = +3/2$.

$$
\begin{array}{||c|ccccc|c||}
\hline \hline
E_{(L)}=E_{(R)}
&|N_{(L)}\rangle& {\stackrel{A^{(3/2)}
}{\longrightarrow}} &|N_{(R)}\rangle= |N_{(\tilde{L})}\rangle&
{\stackrel{A^{(-5/2)} }{\longrightarrow}}
&|N_{(\tilde{R})}\rangle&
E_{(\tilde{L})}=E_{(\tilde{R})}
\\
 \hline
\hline \vdots&\vdots&&\vdots&&\vdots&\vdots
\\
 8&{\cal L}^{(3/2)}_{2}{}&\rightarrow&{\cal L}^{(5/2)}_{1}{}&
\rightarrow&{\cal L}^{(3/2)}_{1}{}&14
 \\
 6&{\cal L}^{(-3/2)}_{3}{}&\rightarrow&{\cal L}^{(-5/2)}_{3}{}&
\rightarrow&{\cal L}^{(-3/2)}_{2}{}&12
 \\
 4&{\cal L}^{(3/2)}_{1}{}&\rightarrow&{\cal L}^{(5/2)}_{0}{}&
\rightarrow&{\cal L}^{(3/2)}_{0}{}&10
 \\
 2&{\cal L}^{(-3/2)}_{2}{}&\rightarrow&{\cal L}^{(-5/2)}_{2}{}&
\rightarrow&{\cal L}^{(-3/2)}_{1}{}&8
 \\
 0&{\cal L}^{(3/2)}_{0}{}&\rightarrow&0&\rightarrow&-&6
 \\
 -2&{\cal L}^{(-3/2)}_{1}{}&\rightarrow&{\cal L}^{(-5/2)}_{1}{}&
 \rightarrow&{\cal L}^{(-3/2)}_{0}{}&4
 \\
-4&-&&-&&-&2\\
 -6&{\cal L}^{(-3/2)}_{0}{}&\rightarrow&{\cal L}^{(-5/2)}_{0}{}&
 \rightarrow&0&0
 \\
 \hline \hline
\ea $$

\end{document}